\newcommand{\nrTotalTerms}[0]{12,339}

\newcommand{\nrArchitectureSpecificBuiltins}[0]{5,480}
\newcommand{\nrArchitectureIndependentBuiltins}[0]{560}
\newcommand{\nrGCCInternalBuiltins}[0]{6,299}

\newcommand{\nrDefBuiltins}[0]{6,040}
\newcommand{\nrUsedBuiltins}[0]{3,083}
\newcommand{\nrProjectsUnfiltered}[0]{4,998}
\newcommand{\nrProjects}[0]{4,913}
\newcommand{\nrFromStars}[0]{80}

\newcommand{\totalMLoc}[0]{1,124}

\newcommand{\avgNrBuiltinsPerProject}[0]{173}
\newcommand{\medianNrBuiltinsPerProject}[0]{9}
\newcommand{\medianNrMachineSpecificBuiltinsPerProject}[0]{69}
\newcommand{\medianNrMachineIndependentBuiltinsPerProject}[0]{7}
\newcommand{\kBuiltins}[0]{319}
\newcommand{\kProjectUniqueBuiltins}[0]{30}
\newcommand{\nrProjectsWithBuiltins}[0]{1,842}
\newcommand{\percentageProjectsWithBuiltins}[0]{37\%}
\newcommand{\percentageRemovedByDoubleQuotes}[0]{2\%}

\newcommand{\percentageBuiltinFilteredByComments}[0]{1\%}
\newcommand{\percentageFilteredByCompilerDirectories}[0]{45\%}

\newcommand{\percentageFilteredByBlacklist}[0]{1\%}
\newcommand{\percentageFilteredBuiltins}[0]{48\%}
\newcommand{\kBuiltinsUnfiltered}[0]{659}
\newcommand{\nrBlacklist}[0]{1,272}

\newcommand{\avgNumberUniqueBuiltinsPerProject}[0]{17}
\newcommand{\medianNumberUniqueBuiltinsPerProject}[0]{4}
\newcommand{\medianNumberMachineSpecificUniqueBuiltinsPerProject}[0]{17}
\newcommand{\medianNumberMachineIndependentUniqueBuiltinsPerProject}[0]{3}
\newcommand{\medianBuiltinEveryXLOC}[0]{5,741}
\newcommand{\avgBuiltinEveryXLOC}[0]{20417}
\newcommand{\minBuiltinEveryXLOC}[0]{7}
\newcommand{\maxBuiltinEveryXLOC}[0]{1,680,582}
\newcommand{\nrProjectsWithMachineIndependentBuiltins}[0]{1,776}
\newcommand{\nrProjectsWithMachineSpecificBuiltins}[0]{422}
\newcommand{\kMachineSpecificBuiltinsTotal}[0]{213}
\newcommand{\kMachineIndependentBuiltinsTotal}[0]{85}
\newcommand{\topOtherBuiltins}[0]{21}
\newcommand{\nrOtherBuiltins}[0]{68}
\newcommand{\topSyncBuiltins}[0]{11}
\newcommand{\topAtomicBuiltins}[0]{7}
\newcommand{\topOtherLibcBuiltins}[0]{4}
\newcommand{\topInternalBuiltins}[0]{4}
\newcommand{\topAddressBuiltins}[0]{3}

\newcommand{\topMachineSpecificBuiltins}[0]{44}
\newcommand{\topPowerpcAltivecBuiltins}[0]{17}
\newcommand{\topArmCBuiltins}[0]{25}

\newcommand{\numberDefBuiltinsNotUsed}[0]{3,033}

\newcommand{\percentageDefBuiltinsNotUsed}[0]{50\%}
\newcommand{\nrMachineSpecificBuiltinsUsed}[0]{2,627}
\newcommand{\nrMachineIndependentBuiltinsUsed}[0]{379}

\newcommand{\percentageUnusedMachineSpecificBuiltins}[0]{52\%}
\newcommand{\percentageUnusedMachineIndependetBuiltins}[0]{32\%}

\newcommand{\classificationsNrAgreed}[0]{36\%}
\newcommand{\classificationsNrTwoAgreed}[0]{46\%}
\newcommand{\classificationsAllDisagreed}[0]{18\%}

\newcommand{\nrUnusedbounds}[0]{11}

\newcommand{\nrUnusedcilk}[0]{13}

\newcommand{\percentageUnusedotherlibc}[0]{37\%}

\newcommand{\medianmostlyincreasing}[0]{18}
\newcommand{\medianmostlystable}[0]{10}
\newcommand{\medianstablethenincreasing}[0]{14}
\newcommand{\medianincreasingthendecreasing}[0]{16}
\newcommand{\mediannoclearpattern}[0]{14}
\newcommand{\medianincreasingthenstable}[0]{12}

\newcommand{\medianspikethenstable}[0]{8}

\newcommand{\projectstatstable}[0]{
\captionof{table}{Overview of the projects obtained (after filtering); the first commit in 1984 stems from a project that was converted from another version-control system.}
\begin{tabular}{l l l l l}
\toprule{}
Metric & Minimum & Maximum & Average & Median \\
\midrule{}%
C LOC & 100 & 37M & 228k & 10k \\
\# commits & 1 & 668k & 4872 & 1147 \\
\# committers & 1 & 17k & 120 & 54 \\
first commit & 1984-02-21 & 2017-11-06 & - & 2011-04-12 \\
last commit & 2003-12-08 & 2017-11-24 & - & 2017-11-07 \\
\bottomrule{}
\end{tabular}
\label{tbl:projectstatstable}}

\newcommand{\smallbuiltintable}[0]{
\captionof{table}{The 10 most frequent builtins.}
\begin{tabular}{l l l}
\toprule{}
builtin & category & projects\\
\midrule{}%
\_\_builtin\_expect & other (compiler interaction) & 890 / 48.3\%\\
\_\_builtin\_clz & other (bitwise operation) & 536 / 29.1\%\\
\_\_builtin\_bswap32 & other (bitwise operation) & 483 / 26.2\%\\
\_\_builtin\_constant\_p & other (compiler interaction) & 430 / 23.3\%\\
\_\_builtin\_alloca & other (stack allocation) & 373 / 20.2\%\\
\_\_sync\_synchronize & sync & 356 / 19.3\%\\
\_\_builtin\_bswap64 & other (bitwise operation) & 347 / 18.8\%\\
\_\_sync\_fetch\_and\_add & sync & 332 / 18.0\%\\
\_\_builtin\_ctz & other (bitwise operation) & 324 / 17.6\%\\
\_\_builtin\_bswap16 & other (bitwise operation) & 304 / 16.5\%\\
\bottomrule{}
\end{tabular}
\label{tbl:smallbuiltintable}}

\newcommand{\percentageincreasing}[0]{39\%}
\newcommand{\nrincreasingthendecreasing}[0]{93}
\newcommand{\percentageincreasingthendecreasing}[0]{14\%}
\newcommand{\nrincreasingthenstable}[0]{140}
\newcommand{\percentageincreasingthenstable}[0]{21\%}

\newcommand{\nrmostlyincreasing}[0]{250}
\newcommand{\percentagemostlyincreasing}[0]{37\%}
\newcommand{\nrmostlystable}[0]{6}
\newcommand{\percentagemostlystable}[0]{1\%}
\newcommand{\nrnoclearpattern}[0]{147}
\newcommand{\percentagenoclearpattern}[0]{22\%}
\newcommand{\nrnotautomatic}[0]{677}
\newcommand{\percentagenotautomatic}[0]{37\%}
\newcommand{\nrspikethenstable}[0]{24}
\newcommand{\percentagespikethenstable}[0]{4\%}
\newcommand{\nrstablethenincreasing}[0]{17}
\newcommand{\percentagestablethenincreasing}[0]{3\%}

\newcommand{\percentagestagnant}[0]{25\%}

\documentclass[sigconf,screen,authorversion]{acmart} 

\setcopyright{acmlicensed}
\acmPrice{15.00}
\acmDOI{10.1145/3338906.3338907}
\acmYear{2019}
\copyrightyear{2019}
\acmISBN{978-1-4503-5572-8/19/08}
\acmConference[ESEC/FSE '19]{Proceedings of the 27th ACM Joint European Software Engineering Conference and Symposium on the Foundations of Software Engineering}{August 26--30, 2019}{Tallinn, Estonia}
\acmBooktitle{Proceedings of the 27th ACM Joint European Software Engineering Conference and Symposium on the Foundations of Software Engineering (ESEC/FSE '19), August 26--30, 2019, Tallinn, Estonia}

\usepackage{hyperref}
\usepackage{booktabs}
\usepackage{tcolorbox}
\usepackage{tabularx}
\usepackage{color}
\usepackage{balance}
\usepackage{listings}
\usepackage[skip=3pt]{caption}
\usepackage{paralist}

\renewcommand{\paragraph}[1]{\textbf{#1}}

\newcommand\code[1]{\texttt{#1}}
\newcommand\quotecommit[1]{\emph{``#1''}}
\newcommand\ninetyninepercprojectsapproxnumberbuiltins{1,600} 
\newcommand\tenbuiltinssupportapproxpercprojects{30\%} 
\newcommand\halfprojectnumberbuiltins{32} 
\newcommand\approxnumberusedbuiltins{3,000} 
\newcommand{\RNum}[1]{\uppercase\expandafter{\romannumeral #1\relax}}

\usepackage{xprintlen}
\usepackage{multirow}

\usepackage[utf8]{inputenc}
\begin{document}
\author{Manuel Rigger}
\affiliation{
  \institution{Johannes Kepler University Linz}
  \country{Austria}
}
\email{manuel.rigger@jku.at}

\author{Stefan Marr}
\affiliation{
  \institution{University of Kent}
  \country{United Kingdom}
}
\email{s.marr@kent.ac.uk}

\author{Bram Adams}
\affiliation{
  \institution{Polytechnique Montr\'eal}
  \country{Canada}
}
\email{bram.adams@polymtl.ca}

\author{Hanspeter Mössenböck}
\affiliation{
   \institution{Johannes Kepler University Linz}
   \country{Austria}
}
\email{hanspeter.moessenboeck@jku.at}

\title{Understanding GCC Builtins to Develop Better Tools}

\begin{abstract}
    C programs can use compiler builtins to provide functionality that the C language lacks.
    On Linux, GCC provides several thousands of builtins that are also supported by other mature compilers, such as Clang and ICC.
    Maintainers of other tools lack guidance on whether and which builtins should be implemented to support popular projects.
    To assist tool developers who want to support GCC builtins, we analyzed builtin use in \nrProjects{} C projects from GitHub.
    We found that \percentageProjectsWithBuiltins{} of these projects relied on at least one builtin.
    Supporting an increasing proportion of projects requires support of an exponentially increasing number of builtins; however, implementing only 10 builtins already covers over \tenbuiltinssupportapproxpercprojects{} of the projects.
    Since we found that many builtins in our corpus remained unused, the effort needed to support 90\% of the projects is moderate, requiring about 110 builtins to be implemented.
    For each project, we analyzed the evolution of builtin use over time and found that the majority of projects mostly added builtins.
    This suggests that builtins are not a legacy feature and must be supported in future tools.
    Systematic testing of builtin support in existing tools revealed that many lacked support for builtins either partially or completely; we also discovered incorrect implementations in various tools, including the formally verified CompCert compiler.
\end{abstract}



\maketitle

\section{Introduction}
Most C programs consist not only of C code, but also of other elements, such as preprocessor directives, freestanding assembly code files, inline assembly, compiler pragmas, and compiler builtins.
While recent studies have highlighted the role of linker scripts~\cite{linkers} and inline assembly~\cite{inlineassembly}, compiler builtins have so far attracted little attention.
Builtins resemble functions or macros; however, they are not provided by libc, but are directly implemented in the compiler.
The following code fragment shows the usage of a GCC builtin that returns the number of leading zeros in an integer's binary representation:
\begin{lstlisting}[language=C, label={leadingzeroes}]
    int leading_zeroes = __builtin_clz(INT_MAX); // returns 1
\end{lstlisting}
On Linux, we observed that GCC builtins are widely used and are also supported by other mature compilers, such as Clang~\cite{clanggccsupport} and ICC~\cite{iccgccsupport}.
For developers working on tools that process C code, such as compilers as well as static and dynamic analysis tools, implementation and maintenance of GCC builtins is a large effort, as we identified a total number of \nrTotalTerms{} GCC builtins, all of which are potentially used by projects and thus need to be supported.
Hence, to assist developers of tools that process C code, the goal of this study was to investigate the use of builtins and how current tools support them.
To this end, we analyzed the builtin use of \nrProjects{} projects from GitHub and implemented a builtin test suite, which we used to test popular tools employed by C developers.

By combining quantitative and qualitative analyses, we answer the following research questions (RQs):

\emph{RQ1: How frequently do projects use builtins?}
Knowing the prevalence of builtins helps tool writers to judge the importance of implementing support for them.
We hypothesized that builtins are used by many projects, and that any program that processes C code will therefore encounter them, yet---similar to inline assembly~\cite{inlineassembly}---we expected that they are used in only a few source-code locations.

\emph{RQ2: For what purposes are builtins used?}
Knowing the primary use cases for builtins helps tool developers to judge whether their tools can support them.
For example, static analysis tools might lack support for multithreading and hence be unable to deal with atomic builtins used for synchronization.

\emph{RQ3: How many builtins must be implemented to support most projects?}
Tool authors who have decided to support GCC builtins would find it helpful to know the implementation order that would maximize the number of projects supported.

\emph{RQ4: How does builtin usage develop over time?}
Understanding the usage of builtins over time could tell us whether projects continue to add builtins or remove them.
If builtins were a legacy feature of compilers that projects sought to remove, the incentive of tool developers to implement them would be low.

\emph{RQ5: How well do tools support builtins?}
To determine the room for improvement in tools, we examined how well existing tools support builtins.
Our assumption was that state-of-the-art compilers such as GCC, Clang, and ICC provide full support, while other tools provide partial or no support.

We found the following:
\begin{compactitem}
\item \nrTotalTerms{} GCC builtins exist, but only \nrUsedBuiltins{} were used in our corpus of projects;
\item \percentageProjectsWithBuiltins{} of the projects used builtins. 
\item Projects primarily used architecture-independent builtins, for example, to interact with the compiler, for bit-level operations, and for atomic operations. However, when a project would use an ar\-chi\-tec\-ture-specific builtin, it would often be used many times in the same project.
\item While mature compilers seem to provide full support for builtins, most other tools lack some builtins or have some implemented incorrectly. Notably, we found two incorrectly implemented GCC builtins in an unverified part of the formally verified CompCert compiler.
\item The effort of supporting a specific number of projects rises exponentially; for example, to support half of the projects only \halfprojectnumberbuiltins{} builtins are needed.
  Supporting 99\% of the projects, however, requires about \ninetyninepercprojectsapproxnumberbuiltins{} builtins.
\item Over time, most of the projects increasingly used builtins; nevertheless, a number of projects removed builtin uses to reduce maintenance effort.
\end{compactitem}

Our results are expected to help tool developers in prioritizing implementation effort, maintenance, and optimization of builtins.
Thus, this study facilitates the development of compilers such as GCC, Clang~\cite{llvm}, ICC, and the formally verified CompCert compiler~\cite{compcert1,compcert2}; of static-analysis tools such as the Clang Static Analyzer~\cite{clangstaticanalyzer}, splint~\cite{splint,lclint}, Frama-C~\cite{framacinline}, and uno~\cite{uno}; of semantic models for C~\cite{depths,typedc11semantics}; and of alternative execution environments and bug-finding tools such as KLEE~\cite{cadar_klee_2008}, Sulong~\cite{native-sulong,asplos}, the LLVM sanitizers~\cite{msan,asan}, and SoftBound~\cite{softbound,cets}.
For reproducibility and verifiability, we provide the database with GCC builtin usage, test suite, tools used for the analysis, and a record of the manual decisions on \emph{\url{https://github.com/jku-ssw/gcc-builtin-study}}.

\section{Methodology}

To answer our research questions, we analyzed builtin use in a large number of C projects and populated a SQLite3 database with the extracted data.
As detailed below, we downloaded and filtered projects from GitHub on which we performed a textual search for the names of the GCC builtins.
To identify the builtin names, we extracted them from the official documentation and from the GCC source code.
To exclude false-positive identifications of builtin use,
we applied heuristics such as excluding builtin names inside string literals and comments.

\paragraph{Selecting the projects.}
We analyzed projects from the popular GitHub code-hosting service.
Similar to other large empirical studies~\cite{qiu:javafeatures,wu:concurrencyconstructs,Casalnuovo:asserts}, we selected projects based on popularity, specifically the number of GitHub stars~\cite{githubpopularity}.
To obtain about 5,000 projects, we downloaded projects down to \nrFromStars{} stars.
This cutoff point was sufficiently large to prevent the inclusion of personal projects, homework assignments, and forks~\cite{perils}.
In total, we downloaded \nrProjectsUnfiltered{} GitHub projects that contained in total \totalMLoc{} million lines of C code.
This strategy allowed us to obtain a diverse set of projects (see Table~\ref{tbl:projectstatstable}).
To provide further evidence for the diversity of the projects, we computed Nagappan et al.'s coverage score~\cite{Nagappan:diversity}. For this, we used the manually validated GitHub project metadata of Munaiah et al.'s RepoReapers data set~\cite{Munaiah2017}, which contained 2,329 of our studied projects. 
This subset alone obtained a coverage score of 0.966 with respect to the universe of 145,355 C projects in the RepoReapers data set. This indicates that our project sample is both representative and diverse.

\begin{table}[!htbp]
    \footnotesize
    \projectstatstable{}
\end{table}

\paragraph{Filtering the projects.}
From the downloaded projects, we selected \nrProjects{} by filtering out those that did not meet our needs.
First, we filtered out all projects that had fewer than 100 LOC, as we considered them too small to constitute C projects.
GCC, forks of GCC\footnote{The projects filtered out included the GCC fork for the Xtensa processor (\url{https://github.com/jcmvbkbc/gcc-xtensa}), and a fork that is based on GCC to dump an XML description of C++ code (\url{https://github.com/gccxml/gccxml}).}, and other C/C++ compilers (such as ROSE~\cite{rose}) implement the GCC builtins themselves, use them internally, and exercise them in their test suites.
Hence, to avoid a high number of false positives, we excluded these projects; they were easy to identify, as they contained the largest numbers of unique builtins.

\paragraph{Identifying the builtins.}
Next, we identified the names of the available GCC builtins, to then perform a textual search on the GitHub projects.
Identifying the list of names was difficult, since GCC builtins are not described or specified in a coherent manner, as they were added over a period of more than 30 years.
Thus, we investigated both (\RNum{1}) builtins listed in the GCC documentation as well as (\RNum{2}) builtins internal to GCC, which we automatically extracted from GCC's source code (including test cases for builtins).

\paragraph{(\RNum{1}) Builtins from the documentation.}
Initially, we considered only builtins described by the GCC documentation.
The GCC documentation stated that some builtins are internal, which we initially did not want to include as we expected that other projects would not use them.
While extracting the names of architecture-independent builtins worked well, GCC also provides builtins that are specific to an architecture.
For example, \code{\_\_builtin\_ia32\_paddq} allows the use of
x86's \code{paddq} instruction.
In some cases, architecture-specific builtins were not described by the documentation, but referred to vendor documentation, for example, the ARM C Language Extensions.
For these builtins, the documentation of GCC version 4.8 contained a list of builtins, which we used instead.
However, for certain special-purpose architectures, obtaining such a list was impractical, for example, for the TILE-Gx and TILEPro processor builtins.
As we expected little influence on the results---overall, architecture-specific builtins were used infrequently (see Section~\ref{sec:usedbuiltins})---we omitted analyzing these special-purpose builtins.
In total, this process yielded \nrDefBuiltins{} builtins, of which \nrArchitectureIndependentBuiltins{} were architecture-independent and \nrArchitectureSpecificBuiltins{} were architecture-specific.

\paragraph{(\RNum{2}) Builtins from the GCC source code}
To verify that we did not omit any commonly used builtins, we searched the projects for strings starting with \code{\_\_builtin\_}.
Since we found that many analyzed projects relied on a small number of GCC's internal (i.e., undocumented) builtins (see below), we assumed that tool developers would also need to support these builtins. Hence, we added them to our search terms by including all additional \code{\_\_builtin\_} functions that we found in the GCC source code and test suite (\nrGCCInternalBuiltins{} additional builtins).
In a number of cases, GCC implemented public builtins using undocumented internal builtins; this was a potential problem in our study, as public and internal builtins would be counted as separate even if they implemented the same semantics.
However, since the number of internal builtins actually used was relatively small, we did not attempt to match public builtins with internal ones in our quantitative analysis.

\paragraph{Searching within the Projects.}
For each analyzed project, we searched all its C files for the names of the \nrTotalTerms{}
builtins described by the GCC documentation or used in the GCC source code.
Note that we considered only occurrences where the builtin name was not a substring of another identifier.
For each builtin that we found, we created a record in our database, thus obtaining \kBuiltinsUnfiltered{}k builtin entries.

\paragraph{Excluding builtin use records.}
We used several strategies to eliminate false positives in the builtin use records.
While investigating the projects with the highest numbers of unique builtins---mostly operating systems---we found that many of them included parts of the source code of Clang or GCC, even though the projects themselves were not compiler projects. Such projects were missed by our prior filtering.
For these projects, we excluded directories whose name started with \code{gcc}, \code{clang} or \code{llvm} (excluding \percentageFilteredByCompilerDirectories{} of our records).

We also excluded builtin occurrences that were enclosed in double quotes,
as this indicates that they are part of a string literal instead of part of the code (excluding \percentageRemovedByDoubleQuotes{} of the records).
To exclude builtins in comments, we did not consider builtins found in lines that started with \code{/*}, \code{*}, or \code{//} (which excluded \percentageBuiltinFilteredByComments{} of the records).

Finally, we manually inspected a number of randomly-selected uses for each distinct builtin, which we used to create a list of \nrBlacklist{} one-line code fragments that indicated false positives (excluding \percentageFilteredByBlacklist{} of the records).
We consider this filtering step optional, since it did not significantly reduce the number of builtin uses.
As part of this process, we detected that builtins not starting with the \code{\_\_builtin\_} prefix (i.e., machine-specific builtins) were likely to cause false positives, which is why we examined such builtin uses in detail.
For example, the TI C6X architecture provided builtins like \code{\_abs}, which often occurred in code that did not use builtins.
As another example, inline assembly with an instruction mnemonic that corresponded to a builtin name often resulted in false positives.
In total, these measures reduced the number of records to \kBuiltins{}k (\percentageFilteredBuiltins{} of the original number).

\section{Results}

\subsection{RQ1: How frequently are builtins used?}
\label{sec:frequency}

To answer RQ1, we considered both duplicate and unique builtin uses per project.
Counting uses---even if they were duplicated within a project---allowed us to measure the overall prevalence of builtin use.
Counting project-unique uses better reflected the implementation effort needed to support a project, because duplicates do not increase the implementation effort.

\begin{figure}[tb]
    \centering
    \includegraphics[width=\columnwidth]{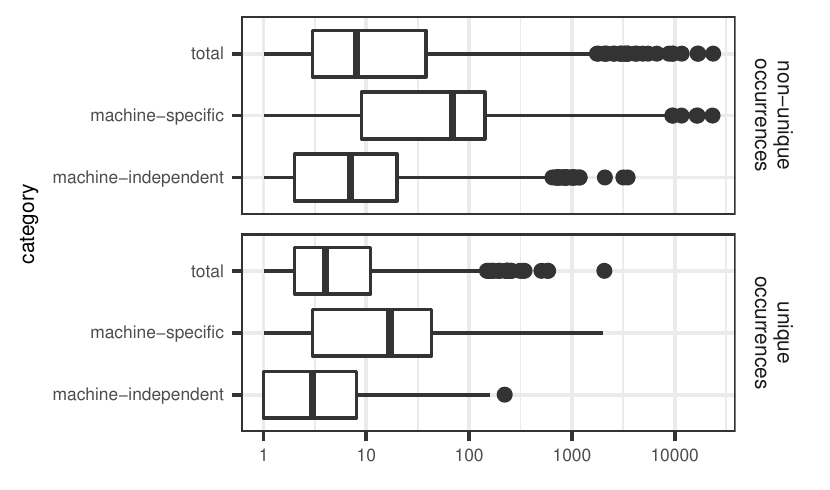}
    \caption{Number of architecture-specific and architecture-independent builtins per project that used builtins of the respective category (logarithmic scale).}
    \label{plot:nruniquefragments}
\end{figure}

\paragraph{Overall use.}
In total, \nrProjectsWithBuiltins{} of the projects (\percentageProjectsWithBuiltins{} of all projects) used a common subset of \nrUsedBuiltins{} builtins.
The frequency of compiler builtins varied strongly, depending on the project, and ranged from one builtin every \minBuiltinEveryXLOC{} LOC to one every \maxBuiltinEveryXLOC{} LOC.
The median frequency of builtins was one every \medianBuiltinEveryXLOC{} LOC (on average one builtin every \avgBuiltinEveryXLOC{} LOC).
Figure~\ref{plot:nruniquefragments} shows boxplots to illustrate the builtin use by the projects, and breaks their use up into architecture-specific and architecture-independent uses, considering both unique and non-unique builtin occurrences within a project.

\paragraph{Non-unique occurrences.}
The median number of builtin calls in a project that used builtins was \medianNrBuiltinsPerProject{}, the average was \avgNrBuiltinsPerProject{}, indicating that there were outlier projects that used a large number of builtins.
In projects that used builtins, architecture-specific builtins were employed in greater numbers (median = \medianNrMachineSpecificBuiltinsPerProject{}); in contrast, when architecture-independent builtins were used, their numbers were far lower (median = \medianNrMachineIndependentBuiltinsPerProject{}).
However, since use of architecture-specific builtins is limited to fewer projects (see Section~\ref{sec:usedbuiltins}), the overall result is dominated by the ar\-chi\-tec\-ture-independent builtins.
We investigated the 15 projects with the highest numbers of builtins and found that audio/video players and codecs lead the ranking (9/15), followed by operating systems (3/15), a game engine, a software library specialized for ARM processors, and a libc implementation.

\paragraph{Unique occurrences.}
Of the \kBuiltins{}k builtin calls, \kProjectUniqueBuiltins{}k were project-unique; that is, the others were duplicated within a project.
The median number of unique builtins used by projects with builtins was low, with a median of \medianNumberUniqueBuiltinsPerProject{} and an average of \avgNumberUniqueBuiltinsPerProject{}.
As with non-unique builtins, projects that used architecture-specific builtins had more such builtins (median = \medianNumberMachineSpecificUniqueBuiltinsPerProject{}) than projects that used architecture-independent builtins (median = \medianNumberMachineIndependentUniqueBuiltinsPerProject{}).
The projects that used the largest number of unique builtins were, again, in most cases audio/video players and codecs (6/15).
However, operating systems (2/15), game engines (2/15), language implementations and compilers (2/15), a messenger, and an image codec also ranked among the top 15.

\paragraph{Reoccurring files.}
We observed that files with particular names, primarily header files, were more likely to contain calls to builtins.
One reason for this was that, consistent with findings by Lopes et al.~\cite{duplicates}, files were copied from other projects.
The majority of these files originated either from the GNU C library \code{glibc} or from Linux-based operating systems.
While they were used primarily in operating system implementations, they were also copied to projects with application code.
As another example, the frequently used \code{sqlite3.c} and \code{SDL\_stdinc.h} files even contained the projects' names as part of the file name: SQLite is a popular database, and SDL a commonly used media library.
In other cases, duplicate file names indicated the use case for the builtin use.
For example, builtin-based atomicity support was often implemented in files named \code{atomic.h}, and math builtins were used in files named \code{math.h}.

\paragraph{Discussion.}We found that \nrTotalTerms{} GCC builtins exist that tool developers potentially need to consider, but that only about \approxnumberusedbuiltins{} of these are used.
Although a builtin is typically found only once every \medianBuiltinEveryXLOC{} lines of C code, \percentageProjectsWithBuiltins{} of all popular projects rely on compiler builtins, thus strongly incentivizing their implementation in analysis and other tools.

\subsection{RQ2: For what purposes are builtins used?}
\label{sec:usedbuiltins}

\begin{figure*}[tb]
    \centering
    \includegraphics[width=\textwidth]{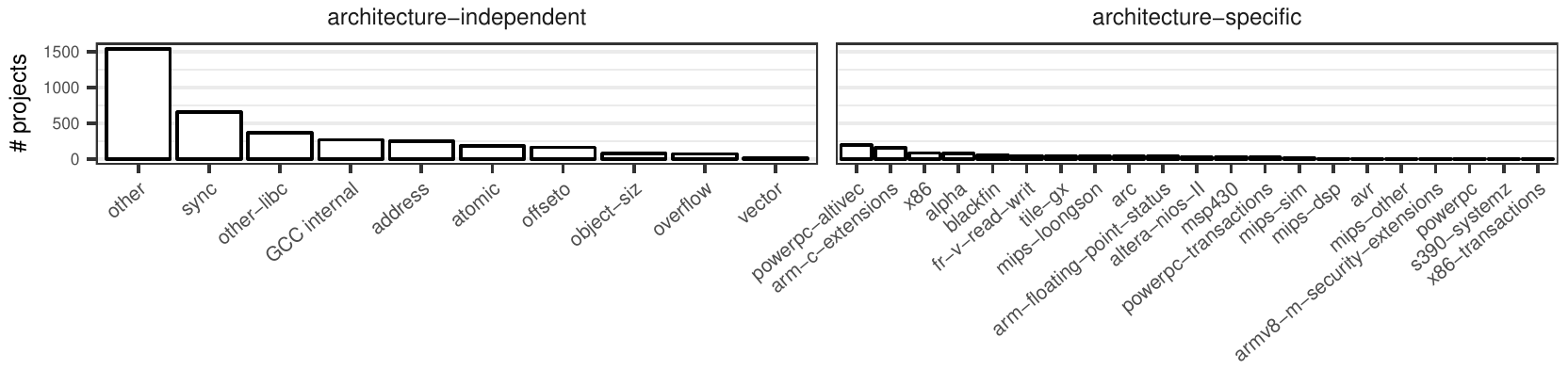}
    \caption{The number of projects that rely on architecture-independent and architecture-specific builtins.}
    \label{img:builtinsbarplots}
\end{figure*}

To identify the purpose for which builtins are used, we explored their usage at different levels of granularity:
First, we examined in detail the usage of architecture-independent builtins and of architecture-specific builtins, as summarized in Figure~\ref{img:builtinsbarplots}.
Then, we analyzed builtins that remained unused in our corpus.

\paragraph{Architecture-specific and -independent builtins.}
The GCC documentation categorizes builtins into architecture-specific and architecture-independent ones, which we used as a basis for discussion.
While \nrProjectsWithMachineIndependentBuiltins{} projects used at least one architecture-independent builtin, we found architecture-specific builtins in only \nrProjectsWithMachineSpecificBuiltins{} projects.
That architecture-independent builtins are more common across project was unexpected, since we found only \kMachineIndependentBuiltinsTotal{}k architecture-independent builtin uses, but \kMachineSpecificBuiltinsTotal{}k architecture-specific ones.
However, as discussed in Section~\ref{sec:frequency}, a project using architecture-specific builtins is likely to use more such builtins than projects that use architecture-in\-de\-pen\-dent builtins.

\paragraph{``Other'' builtins.}
The builtin category ``other'', which contained miscellaneous builtins, was the most common category of GCC builtins, even though it comprised only \nrOtherBuiltins{} builtins---\topOtherBuiltins{} of which were among the 50 most frequently used.
Since these builtins were the most common, we further analyzed their use, and classified them into the following subcategories: (\RNum{1}) direct compiler interaction, (\RNum{2}) bit and byte operations, (\RNum{3}) special floating-point values, and (\RNum{4}) dynamic stack allocation. 

\subparagraph{(\RNum{1}) Direct compiler interaction.}
\sloppy
These builtins allow direct interaction with the compiler, for example, to improve performance; the most frequently used builtin was \code{\_\_builtin\_expect}, which communicates expected branch probabilities to the compiler, which can exploit this information for optimization.
The \code{\_\_builtin\_unreachable} builtin can be used to silence warnings by informing the compiler that code is unreachable, which is useful when the compiler cannot deduce this.
Some of the builtins in this subcategory can also be used for metaprogramming; the \code{\_\_builtin\_constant\_p} builtin is resolved at compile time and allows programmers to query whether a pointer is known by the compiler to be constant.
As another example, \code{\_\_builtin\_types\_compatible\_p} queries whether two input types passed to the builtin are the same.
Plain C does not offer similar functionality.

\subparagraph{(\RNum{2}) Bit and byte operations.}
These builtins process integers at the level of bits and bytes.
The second-most frequently used builtin was \code{\_\_builtin\_clz}, which counts the leading zeroes in an \code{unsigned int}; its variants for other data types also ranked among the most commonly used builtins overall.
Similarly frequent were builtins for computing the position of the least significant one-bit, for counting the number of one-bits in an integer, and for reversing the bytes of an integer.
We believe that these builtins were used for convenience and performance optimizations, as the same functionality could be implemented in plain C.

\subparagraph{(\RNum{3}) Special floating-point values.}
These builtins generate special values for various floating-point types.
For example, the \code{\_\_builtin\_inf} builtin generates a positive infinity \code{double} value.
As another example, \code{\_\_builtin\_nan} returns a not-a-number value.
Recent C standards specify macros and functions for obtaining such values.

\subparagraph{(\RNum{4}) Dynamic stack allocation.}
The \code{\_\_builtin\_alloca} builtin allocates the specified number of bytes of stack memory.
Since C99, variable length arrays have offered a similar functionality, as the size of an allocated array can depend on a run-time value.

\paragraph{Synchronization and atomics.}
After ``other'', the next common builtin category was synchronization (``sync'') with \topSyncBuiltins{} of the 50 most common builtins.
In this category, the most frequently used builtin was \code{\_\_sync\_synchronize}, which issues a full memory barrier to restrict the order of execution in out-of-order CPUs.
Builtins for atomically executing operations were also common (e.g., \code{\_\_sync\_fetch\_and\_add}).
These builtins were designed for the Intel Itanium ABI and were deprecated in favor of the builtins contained in the ``atomic'' category.
The builtins in the ``atomic'' category additionally allow specifying the memory order of the operation, but were not that frequently used; nevertheless \topAtomicBuiltins{} builtins of this category ranked among the 100 most common builtins.
Note that C11 introduced synchronization primitives, which are alternatives to these builtins.

\paragraph{Libc functions.}
GCC provides builtins for many functions of the standard C library---\topOtherLibcBuiltins{} such builtins were amongst the 100 most common builtins.
An example is \code{\_\_builtin\_memcpy}, which implements the semantics of \code{memcpy}.
The builtin version of the libc function is useful when compiling a program assuming a C dialect in which a function is not yet available; for example, when compiling under the C90 standard (\code{-std=c90}), the newer C99 function \code{log2} cannot be used; however, the prefixed version \code{\_\_builtin\_log2} can still be used.
Furthermore, they enable bare-metal programs, which are compiled \emph{freestanding} and therefore do not have access to libc functions, unless they use compiler builtins.

\paragraph{GCC internal functions.}
Several builtins were used by projects although they were not documented---\topInternalBuiltins{} ranked among the top 100 frequently used builtins.
These most frequently used builtins, namely \code{\_\_builtin\_va\_start}, \code{\_\_builtin\_va\_end}, \code{\_\_builtin\_va\_arg}, and \code{\_\_builtin\_va\_copy}, were used exclusively to implement the vararg macros of the C standard.

\paragraph{Function return address and offsetof.}
The ``introspection'' category---with \topAddressBuiltins{} of the top 100 builtins---enables programmers to query (\RNum{1}) the address to which a function returns and (\RNum{2}) the address of the current frame (i.e., the area where local variables are stored).
To this end, GCC provides \code{\_\_builtin\_return\_address}, \code{\_\_builtin\_frame\_address} and other builtins.
Another, similar category is ``offsetof'' with a single builtin \code{\_\_builtin\_offsetof}, which was one of the top 100 builtins.
It determines the offset of a struct or array member from the start address of the struct or array.

\paragraph{Object size and safe integer arithmetics.}
The builtin \code{\_\_builtin\_object\_size} in the ``object-size'' category enables programmers to query the size of an object, which is useful when implementing bounds checks.
To implement this builtin, GCC relies on static analysis to determine the size of an object where possible.
The ``overflow'' category---of which no builtin ranked among the top 100---provides wraparound semantics for overflow in signed-integer operations (e.g., \code{\_\_builtin\_add\_overflow} for addition), which would otherwise induce undefined behavior in C~\cite{intoverflow}.

\paragraph{Usage of architecture-specific builtins.}
\sloppy
Of the 100 most-frequent builtins, \topMachineSpecificBuiltins{} were specific to an architecture.
Most frequent were the builtins for the PowerPC family---\topPowerpcAltivecBuiltins{} of which were among the top 100 builtins.
The most frequent PowerPC builtins were those implementing vector operations such as \code{vec\_perm}, which implements a vector permutation.
The second category were ARM C NEON extensions---\topArmCBuiltins{} of which were among the top 100 builtins---that also implement vector operations.
On x86, which ranked next, the most common builtin was \code{\_\_builtin\_cpu\_supports} followed by \code{\_\_builtin\_cpu\_init}, which allow programmers to query the availability of CPU features such as SIMD support.
In x86-64 inline assembly, the equivalent \code{cpuid} instruction ranked among the most commonly used instructions~\cite{inlineassembly}.
Other x86 builtins were quite diverse and less frequent.
For brevity, the less frequently used architecture-specific builtin categories are omitted. However, they are included in the full list of commonly used builtins in the online appendix.

\paragraph{Unused builtins.}
To identify unused builtins, we considered only those described in the GCC documentation (i.e., the public ones).
Surprisingly, we found that half of them, namely \numberDefBuiltinsNotUsed{} (\percentageDefBuiltinsNotUsed{}), were not used in our corpus.
The distribution differed between architecture-specific and architecture-independent builtins.
From the architecture-independent builtins, \nrMachineIndependentBuiltinsUsed{} of \nrArchitectureIndependentBuiltins{} were used, which corresponds to \percentageUnusedMachineIndependetBuiltins{} unused builtins.
We characterize these unused builtins below.
From the architecture-specific builtins, only \nrMachineSpecificBuiltinsUsed{} of \nrArchitectureSpecificBuiltins{} builtins were used, which means that more than half of them (\percentageUnusedMachineSpecificBuiltins{}) were not used in any project; this is why we do not characterize them in detail.

We contacted the GCC developers to report our findings~\cite{gccdevs};
they responded that builtins could not be removed from the documentation due to vendor guarantees (for architecture-specific builtins) and because they might still be used in closed-source software or by projects not hosted on GitHub.
While the possibility cannot be excluded that these builtins are used by some projects (or code yet to be written), the possibility that all of them are used is rather low.
Thus, our study provides a first step towards deprecating unused builtins, and removing those builtins from the public documentation that could be considered internal.

\paragraph{Unused architecture-independent builtins.}
None of the projects used any of the \nrUnusedbounds{} bounds-checking builtins for controlling the Intel MPX-based pointer-bounds-checker instrumentation, which is based on a hardware extension in Intel processors.
One reason for this is that they are used by a pass within GCC and have received only little further attention~\cite{rigger2018context}, as Intel MPX-based approaches perform only about as fast as pure software approaches~\cite{intelmpx}.
Four of the object-size-checking builtins were not used, namely a subset of those for printing format strings (e.g., \code{\_\_builtin\_\_\_vfprintf\_chk}).
The builtins of this category were derived from library functions (e.g., \code{memcpy}), but require an additional size argument (e.g., \code{\_\_builtin\_\_\_memcpy\_chk}).
The intended use of these builtins is to prevent buffer overflow attacks, since object accesses that exceed the size of the object can be prevented.
We speculate that these builtins were not frequently used because neither the C language nor builtins provide the functionality to reliably query the size of an object, which would require run-time support~\cite{introspection}.

None of the \nrUnusedcilk{} builtins of the Cilk Plus C/C++ language extensions~\cite{cilkplusplus}, which offer a mechanism for multithreading, were used.
In 2017, Cilk Plus was deprecated, and in November 2017 GCC removed its implementation~\cite{cilkremove}.
Of the prefixed libc functions, \percentageUnusedotherlibc{} were unused.
Most programs are probably compiled in hosted mode, where compilers
can substitute calls to the libc functions with these builtins.
Another reason could be that some of them are used only internally.
Nevertheless, they were documented in the public API.

Of the unused builtins in the ``other'' category, the majority were narrowly specialized builtins such as \code{\_\_builtin\_inffn}, which generates an infinity value for the data type \code{\_Floatn}.
Further, \code{\_\_builtin\_\_\_clear\_cache} for flushing the processor's instruction cache remained unused.
The unused \code{\_\_builtin\_call\_with\_static\_chain} enables calls to languages that expect static chain pointers, such as Go.

\paragraph{Discussion.}
The use cases for builtins were diverse.
The use of GCC builtins was dominated by architecture-independent builtins for direct interaction with the compiler, for bit-and-byte operations, atomic operations, and libc equivalents.
Depending on the tool, different builtin categories could be supported to different degrees; for example, static analysis tools that do not analyze the semantics of multithreaded atomic operations might eschew implementing those.
Architecture-specific builtins were used by fewer projects, but, within these projects, in greater number than architecture-independent builtins.
They were used for SIMD instructions, to determine CPU features, and to access platform-specific registers.

\subsection{RQ3: How many builtins must be implemented to support most projects?}
\label{sec:builtinsupport}

In order to provide tool developers with a recommended implementation order for builtins, we considered two implementation scenarios.
The first scenario considered all builtins as implementation candidates.
The second considered only architecture-independent builtins, which can be relevant when only a subset of architectures is to be supported.
Additionally, we assumed two pragmatic strategies for the order of implementation: an order based on the frequency of builtins, and one based on a greedy algorithm.
Note that this paper assumes equal weights across projects, since weights would have biased the results based on assumptions that might not hold for all tools.

\paragraph{Frequency order.}
Using this strategy, we assumed that the builtins used by the highest number of projects are to be implemented first.
Thus, this strategy follows the order given by Table~\ref{tbl:smallbuiltintable}.
This order is not generally optimal, because it does not take into account that, in order for a project to be supported, all builtins used must be implemented.

\paragraph{Greedy order.}
For rapid experimentation, it can be beneficial to quickly support as many projects as possible.
To this end, we implemented a greedy order where the next builtin to be implemented is selected such that it enables support of the largest number of additional projects.
If no such builtin exists, the next builtin is selected using the frequency order.

\begin{figure}[tb]
    \centering
    \includegraphics[width=\columnwidth]{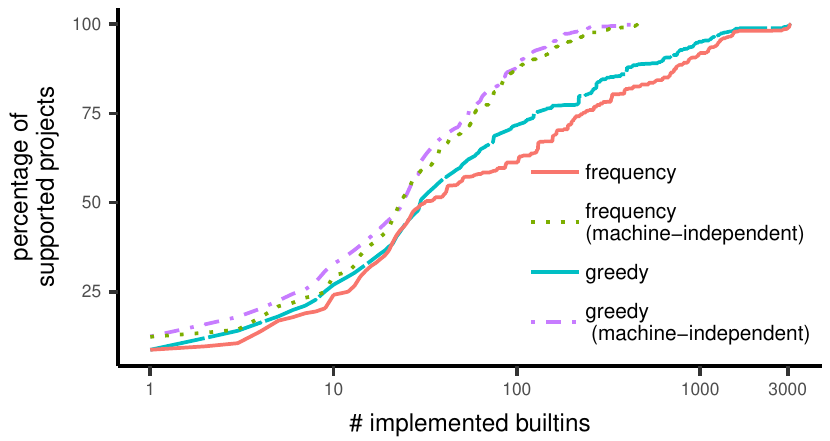}
    \caption{The numbers of builtins needed to support an increasing number of projects increases exponentially; note the exponential x-axis.}
    \label{img:implementationeffort}
\end{figure}

\paragraph{Results.}
Implementing builtins takes an exponential implementation effort in terms of number of builtins that must be implemented to support a specific number of projects (see Figure~\ref{img:implementationeffort}).
The greedy order for implementing builtins performs better than the frequency order, a trend that is more clear-cut when considering all builtins rather than just architecture-independent ones.
To support half of the projects, in both scenarios and using both strategies, no more than \halfprojectnumberbuiltins{} builtins need to be implemented.
Note that these builtins are all architecture-independent ones; this is expected, because, as described in Section~\ref{sec:frequency}, projects rely less frequently on architecture-specific builtins, but
if they do, they use a larger number of such unique builtins.

Supporting 90\% of the projects requires 106 builtins to be implemented for the greedy approach and 112 builtins for the frequency strategy when considering only architecture-independent builtins.
When considering all builtins, more than 850 builtins must be implemented for the frequency strategy, and more than 600 for the greedy strategy.
To support 99\% of the projects, the greedy algorithm is better: when considering only architecture-independent builtins, around 250 instead of 300 builtins must be implemented, compared to \ninetyninepercprojectsapproxnumberbuiltins{} instead of \approxnumberusedbuiltins{} builtins when considering all builtins.
Thus, we suggest that tool developers use a greedy approach when implementing builtins.

\begin{table}
    \footnotesize
    \smallbuiltintable{}
\end{table}

\subsection{RQ4: How does builtin usage develop over time?}

To understand whether builtin usage is an ongoing concern of software projects or just a form of technical debt (introduced temporarily before being removed), we studied the development of builtin usage over time in the projects that used builtins. For this, we analyzed all commits by iterating from the latest commit to the oldest commit---including merge commits (represented by the union of all commits that are merged)---always by following the first parent (i.e., staying on the master branch).
We considered only those projects for further inspection that had at least five commits that introduced or removed calls to builtins, since projects with fewer commits made it difficult to judge a project's development trend.
This left us with \nrnotautomatic{} projects, \percentagenotautomatic{} of the projects for which we processed the builtin history.

\paragraph{Manual inspection methodology.}
We manually classified the \nrnotautomatic{} projects based on their trends of adding and removing builtins.
Since manual classification of trends is partly subjective,
we performed manual classification based on ``negotiated agreement''~\cite{Parnin2017,campbell13}.
Basically, the three authors jointly open-coded the qualitative data sources, arriving at a classification through consensus. Given the lack of pre-existing classifications, such an approach seems justified. In particular, the three authors first independently classified a fixed set of 15\% randomly-selected projects with respect to their builtin trends.
In \classificationsNrAgreed{} of the cases, all three authors agreed on the classification.
In \classificationsNrTwoAgreed{} of the cases, two authors agreed.
In \classificationsAllDisagreed{} of the cases, all authors disagreed.
Subsequently, the three authors discussed diverging classifications and came to a consensus for each of them.
As with other studies~\cite{15-percent,10-15-percent}, this initial classification served as a ``calibration phase'' for a single author to classify the remaining trends.

\paragraph{Classification Results.}
The final classification consisted of four main categories of trends (see Table~\ref{tbl:trendtable}).
Most prevalent was the \underline{\emph{Increasing}} trend, which we assigned to projects that mostly added builtins (\percentageincreasing{}).
The majority of those showed a clear increasing trend (\percentagemostlyincreasing{}), while few had an initial stable period that was followed by an increasing trend (\percentagestablethenincreasing{}).
The second most common trend was the \underline{\emph{Stagnant}} trend (\percentagestagnant{}) for those projects that initially had builtin-related commits, but then did not show any or few further changes to the usage of builtins.
Most \emph{Stagnant} projects initially added builtins, then became stagnant (\percentageincreasingthenstable{}).
Others initially added builtin uses, but then removed all or many of them shortly afterwards---a development to which we refer as a spike---and subsequently showed none or few further changes (\percentagespikethenstable{}).
A low number of \emph{Stagnant} projects exhibited a mostly stable trend overall (\percentagemostlystable{}).
We assigned the \underline{\emph{Decreasing}} trend to projects that initially had an increasing trend followed by a decreasing trend (i.e., the removal of builtin uses, \percentageincreasingthendecreasing{}).
Finally, we assigned the \underline{\emph{Inconclusive}} trend to projects for which we could not clearly assign a trend (e.g., because they exhibited a combination of trends, \percentagenoclearpattern{}).

\begin{table}
    \footnotesize
    \captionof{table}{Builtin trends in projects.}
    \begin{tabular}{p{1.2cm} l r r p{1cm}}
    \toprule{}
    trend & classification & \multicolumn{2}{r}{\#/\% projects} & median commits\\
\midrule{}%
\multirow{2}{*}{Increasing}    & mostly increasing & \nrmostlyincreasing{} & \percentagemostlyincreasing{} & \medianmostlyincreasing{} \\
    & stable, then increasing & \nrstablethenincreasing{} & \percentagestablethenincreasing{} & \medianstablethenincreasing{} \\
\midrule{}%
\multirow{3}{*}{Stagnant}    & increasing, then stable & \nrincreasingthenstable{} & \percentageincreasingthenstable{} & \medianincreasingthenstable{} \\
    & spike, then stable & \nrspikethenstable{} & \percentagespikethenstable{} & \medianspikethenstable{} \\
    & mostly stable & \nrmostlystable{} & \percentagemostlystable{} & \medianmostlystable{} \\

\midrule{}
Decreasing    & & \nrincreasingthendecreasing{} & \percentageincreasingthendecreasing{} & \medianincreasingthendecreasing{} \\
\midrule{}%
    Inconclusive & & \nrnoclearpattern{} & \percentagenoclearpattern{} & \mediannoclearpattern{} \\
    \bottomrule{}
    \end{tabular}
    \label{tbl:trendtable}
\end{table}

\paragraph{Reasons for builtin additions or removal.}
We attempted to find reasons for changes in the numbers of builtins, for which we analyzed commit messages and commit changes, then identified common cases.

\subparagraph{\indent\emph{Builtin additions.}}
The majority of sharp increases in the number of builtins was caused by the inclusion of third-party libraries that call builtins internally, as indicated by commits such as \quotecommit{update packaged sqlite to 3.8.11.1} or \quotecommit{Added latest stb\_image.}
In some cases, only single existing header files were included,
as indicated by commit messages such as \quotecommit{add atomic.h that wraps GCC atomic operations} or \quotecommit{Copy over stdatomic.h from freebsd.}

Builtins, both architecture-specific and -independent ones, were often used for performance optimizations.
Example architecture-independent optimizations are 
\quotecommit{popcount() optimization for speed} (using \code{\_\_builtin\_popcount}), \quotecommit{Use \_\_builtin\_expect in scanline drawers to help gcc predict branching}, and \quotecommit{A prefetch of status-$>$last\_alloc\_tslot saved 5\%} (using \code{\_\_builtin\_prefetch}).
Examples of architecture-specific builtin commits were \quotecommit{VP9 common for ARMv8 by using NEON intrinsics} and \quotecommit{30\% encoding speedup: use NEON for QuantizeBlock()}.

Builtins were also used when they conveniently provided required functionality in commits such as \quotecommit{bitmap -- Add few helpers for [bit] manipulations}.
They were often used for atomics, as in \quotecommit{GCC 4.1 builtin atomic operations} and \quotecommit{Adding atomic bitwise operations api and rwlocks support}.
They enabled metaprogramming techniques, for example, by enabling macros to handle various data types: \quotecommit{util: Ensure align\_power2() works with things other than uint. This uses a [cascading] set of if (\_\_builtin\_types\_compatible\_p()) statements to pick the correct alignment function tailored to a specific type [...]}.

Finally, builtins were employed to reduce the usage of assembly and inline assembly in commits such as \quotecommit{avoid inline assembly in favor of gcc builtin functions} and \quotecommit{Padlock engine: make it independent of inline assembler.}, or as an alternative to architecture-specific system libraries, such as \quotecommit{alloca fallback for gcc}, which added a use of \code{\_\_builtin\_alloca} when the platform did not provide a header file that implements \code{alloca}.

\subparagraph{\indent\emph{Builtin removals.}}
Removals of third-party libraries accounted for the most significant number of removals of builtins, as indicated by commits such as \quotecommit{Remove thirdparties} or \quotecommit{Removed outdated headers and libraries.}
Individual files or functions that used builtins were removed as side effects of refactoring or cleanup in commits with messages such as \quotecommit{General cleanup of the codebase, remove redundant files.} or \quotecommit{tools: Remove unused code.}
Auto-generated files were removed, for instance, in the commit \quotecommit{Removed getdate.c as it is regenerated from getdate.y}.

A number of removals were related to technical debt~\cite{Cunningham1993}.
Projects removed builtins for old architectures for which they dropped support, for instance, in \quotecommit{avr32: Retire AVR32 for good. AVR32 is gone. [...]}
or \quotecommit{Blackfin: Remove. The architecture is currently unmaintained, remove}.
In other cases, builtins for certain architectures were removed due to their maintenance effort: \quotecommit{Remove support for altivec using gcc builtins, since these keep changing across gcc versions. [...]}.
Uses of builtins were hidden behind a macro, to concentrate their use to a single location in the source code: \quotecommit{Convert remaining \_\_builtin\_expect to likely/unlikely [...]}
(for \code{\_\_builtin\_expect}) and \quotecommit{Use the new sol-atomic.h API instead of directly GCC intrinsics} (for atomic operations).

In other cases, a use of \code{\_\_builtin\_expect} was removed because it did not improve performance: \quotecommit{[...] It had no reliably measurable performance improv[e]ment, at least on an i7 960 and within a microbenchmark.}.

\paragraph{Case study.}
\begin{figure*}[tb]
    \centering
    \includegraphics{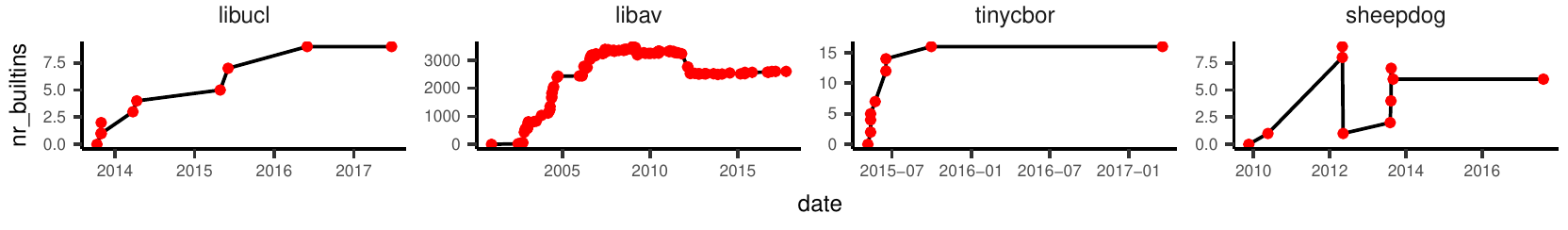}
    \caption{Builtin development in tinycbor, hashcat, libav, and libucl.}
    \label{trends}
\end{figure*}
Finally, we examined the builtin development in four projects whose trends we considered both representative and insightful for our case study (see Figure~\ref{trends}).
First, we selected libucl, a configuration library parser, which is representative of the \emph{Increasing} trend.
Like the majority of projects that we examined, it added a small number of builtin calls for various tasks.
We selected libav, a collection of cross-platform tools to process multimedia formats and protocols, to represent the \emph{Decreasing} category.
As is typical of a media library, it contained a number of builtin-related commits that improved performance by adding calls to architecture-specific builtins, but also systematically removed them to reduce maintenance effort.
We selected tinycbor, a library for encoding and decoding the CBOR format, to represent the \emph{Stagnant} trend.
Specifically, we classified it as \emph{increasing, then stable}.
Finally, we selected sheepdog, a distributed storage system for the QEMU virtual machine, which we classified as \emph{Inconclusive}, due to its ``pit shape'' in the center of the plot.

\paragraph{libucl (Increasing).}
The builtin additions in libucl were in most cases related to hashing.
The first two builtin-related commits of libucl imported a hash algorithm from third-party libraries that used \code{\_\_builtin\_clz} and \code{\_\_builtin\_swap32} in their hashing computations.
Subsequently, a third-party library hashing implementation was replaced with a custom implementation, removing a builtin use.
Subsequent commits were also related to finding better hashing algorithms, resulting in additions of calls to byteswap builtins and checks for SIMD support using \code{\_\_builtin\_cpu\_init} and \code{\_\_builtin\_cpu\_supports}.
Additionally, the library added a reference-counting scheme to free memory when an allocation is no longer referenced, whose implementation depended on atomics.

\paragraph{libav (Decreasing).}
In the first half of libav's development, its use of builtins mainly increased, mostly due to Altivec-specific builtins used to optimize computation-intensive operations, but also due to architecture-specific builtins of other architectures such as PowerPC or ARM.
In a few cases, calls to architecture-independent builtins were added, for example for atomics.
In the second half of the project, refactorings reduced the number of builtin calls.
In 2009, calls to 236 Altivex-specific builtins were removed to reduce technical debt and improve the maintainability of the Snow codec (which was removed in 2012): \quotecommit{Remove AltiVec optimizations for Snow. They are hindering the development of Snow, which is still in flux.}
In 2012, calls to 233 builtins were removed as part of a cleanup that dropped an unused function; in the same year, a library was removed that used 469 builtins.
In 2013, another smaller, but interesting, commit removed calls to 23 Alpha-specific builtins, as the platform was no longer considered important: \quotecommit{Remove all Alpha architecture optimizations. Alpha has been end-of-lifed and no more test machines are available.}

\paragraph{tinycbor (Stagnant)}
The initial commit introduced macros for performance optimizations that used \code{\_\_builtin\_expect} to communicate branch probabilities to the compiler; one of the macros was used to annotate an error handling case as unlikely.
Similarly, the \code{\_\_builtin\_unreachable} builtin was used to annotate the case that should not happen as undefined, allowing the compiler to generate more efficient code.
To support byteswap operations on non-Linux systems, where the \code{endian.h} header file is typically not present, a use of \code{\_\_builtin\_bswap64} was added.
A subsequent commit also introduced byteswap uses for Linux systems, with the commit message stating that it was more efficient and made cross-building the project easier.
The \code{\_\_builtin\_add\_overflow} was added to implement an addition that does not cause undefined behavior on overflow~\cite{intoverflow}.
Three commits that did not change the number of used builtins adjusted the conditions when builtins were used due to portability reasons.
For example, according to the commit messages, \code{builtin\_bswap16} was added with GCC 4.8 and ICC did not support \code{\_\_builtin\_add\_overflow}, making it necessary to check for these cases using macros.
While the last builtin-related commit was in 2015, the project continued to be active until 2017.

\paragraph{sheepdog (Inconclusive)}
In sheepdog, the prominent increase before the pit was caused by a commit that replaced mutex locks by equivalent synchronization builtins, as it was stated to make the execution faster.
The uses were then replaced with calls to an external library that offered equivalent functionality, resulting in the sharp decrease.
Other commits replaced an assembly fragment that obtained the address of the frame pointer with \code{\_\_builtin\_frame\_address} for logging.
The builtin in turn was replaced by invoking gdb to perform this action.
The performance of logging was improved by \code{\_\_builtin\_expect}, which was used to annotate code to assume the standard logging level.
Besides, bit operations were simplified using \code{\_\_builtin\_clzl} and \code{\_\_builtin\_ffsl}.

\paragraph{Discussion.}
We analyzed the development history of builtins in projects and found that many projects mostly
added calls to builtins.
They were added for performance optimizations, atomic implementations, to enable metaprogramming techniques, and others; they were removed, for example, due to their maintenance cost and through refactorings.
Overall, it seems that compiler builtins are not a legacy feature from times when compilers applied less sophisticated optimizations; tool developers must expect that contemporary and future code will use them.

The four representative projects gave insights into how projects added and removed builtin uses.
Like the majority of projects we examined, libucl, tinycbor, and sheepdog had few commits related to architecture-independent builtins.
These builtins were used in various use cases, for instance, to improve the performance of code, to test for CPU features, to implement hash computations, and as a fallback when architecture-specific builtins were missing.
Libav was one of the relatively few projects that had a large number of commits related to architecture-specific builtins, and it reduced their number during code refactorings.
For sheepdog and libav, builtins were also removed to reduce technical debt; in sheepdog, builtins were replaced by using an external library instead, and in libav they were removed since an outdated architecture was no longer supported.

\subsection{RQ5: How well do tools support builtins?}
\begin{figure}[tb]
    \centering
    \includegraphics[width=\columnwidth]{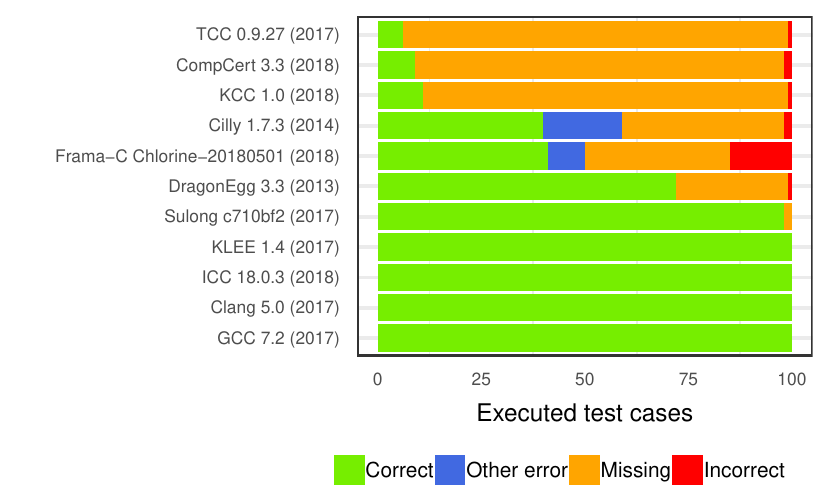}
    \caption{The test-case results before bugs were fixed or builtins implemented. The order of the legend items corresponds to the order of the stacked plots. ``Other errors'' refers to failed test cases unrelated to builtin support.}
    \label{img:tool-evaluation}
\end{figure}

To determine how well current tools support GCC builtins, we manually implemented a builtin test suite for the 100 most commonly used architecture-independent builtins (cf. RQ2), which would support the architecture-independent portion of almost 90\% of the builtin-using projects (see Section~\ref{sec:builtinsupport}).
For each builtin, we used its documentation to determine both typical inputs and corner cases, then wrote test cases for them.
As tools to be tested, we selected popular and widely used mature compilers, special-purpose compilers, source-to-source translators, alternative execution environments, and static analysis tools.
Figure~\ref{img:tool-evaluation} shows the results.

\paragraph{Mature compilers.}
We tested the most widely used open-source compilers on Linux, GCC and Clang~\cite{llvm}, as well as the commercial ICC.
They all executed the test cases successfully.

\paragraph{Special-purpose compilers.}
We tested the special-purpose compilers CompCert~\cite{compcert1,compcert2} and TCC.
CompCert is a compiler used in safety-critical applications and has been formally verified to be correct, which, however, excludes its implementation of builtins.
We found that CompCert correctly executed only 9 builtin test cases, supporting 5 out of the 10 most frequently used builtins.
Both \code{\_\_builtin\_clzl} and \code{\_\_builtin\_ctzl} computed an incorrect result for large input values~\cite{compcertissue}.
After reporting the bugs detected by our test suite, they were fixed within a day with the note that ``we need more testing here''.

The TCC compiler is a small compiler developed to compile code quickly.
It successfully ran only six builtin test cases.
While most tests failed with a build error, the \code{\_\_builtin\_types\_compatible\_p} builtin produced an incorrect result when comparing enumerations~\cite{tccissue}.

\paragraph{C front end.}
The C Intermediate Language (CIL)~\cite{cil} is a front end for the C language that facilitates program analysis and transformation.
We tested its driver, called cilly, which can also be used as a drop-in replacement for GCC.
It successfully executed 40 builtin test cases.
The \code{\_\_builtin\_bswap16} and \code{\_\_builtin\_types\_compatible\_p} builtins produced incorrect results~\cite{cilissue}.
Cilly also failed on 34 atomic test cases, on 15 test cases due to a failure to parse a system library, on 5 test cases due to unrecognized builtins, and on 4 test cases due to warnings for the \code{long double} type.

\paragraph{Source-to-source translators.}
We evaluated DragonEgg, which compiles source languages supported by GCC to LLVM IR.
Although it has not been updated for several years, it successfully executed more than two thirds of the test cases.
It failed to translate more recent builtins (e.g., from the ``atomic'' category) that were added to GCC after the last commit in DragonEgg.

\paragraph{Static analysis.}
We tested Frama-C~\cite{Cuoq2012,Kirchner2015}, a static-analysis framework.
By default, it assumes code to be portable, and supports compiler extensions only with an option.
For 41 test cases, Frama-C's analysis did not trigger a warning or error~\cite{framacissue}.
9 test cases failed because its standard library lacked macros for \code{INFINITY} and \code{NAN}, which were used in the test cases.
14 test cases for \code{\_\_sync} builtins were generally supported, but incorrectly implemented for the \text{long} type.
Furthermore, \code{\_\_builtin\_object\_size} referred to an undefined variable in its macro, which resulted in an error.

\paragraph{Alternative execution environments.}
We tested Sulong~\cite{native-sulong,asplos}, an interpreter with dynamic compiler for LLVM-based languages, and KCC~\cite{Ellison2012,undefinedness}, a commercial interpreter for C that was automatically derived from a formal semantics for C and detects Undefined Behavior.
Sulong successfully executed all but two test cases, namely for \code{\_\_builtin\_fabsl} and \code{\_\_builtin\_\_\_clear\_cache}, which were not implemented~\cite{sulongissue}.
Note that we found these errors with a preliminary version of the test suite, and consequently contributed implementations for the two missing builtins.

KCC successfully executed test cases for 10 builtins, but, since it is based on CIL, it had the same error in the implementation of the \code{\_\_builtin\_types\_compatible\_p} builtin.
The KCC developers also mentioned that they have ``recently been trying to add more supports for gnuc builtins.''~\cite{kccissue}.

\paragraph{Symbolic execution engine.}
We tested KLEE~\cite{cadar_klee_2008}, a symbolic execution for LLVM-based languages.
KLEE executed all test cases successfully when executed with concrete inputs.

\paragraph{Discussion.}
Our findings indicate that mature compilers support builtins, which is expected, since many projects rely on them.
However, many other tools lack builtin implementations or have errors in their implementations.
Note that working builtin implementations can typically not be reused by other tools due to their differences in use cases and implementation languages.
For example, while GCC translates builtin usages to efficient machine code in its C/C++ source code, Frama-C abstractly reasons about them using OCaml.
Tools based on existing mature compiler infrastructure---such as KLEE and Sulong, which are based on LLVM---seem to have a better builtin support, partly because some builtins are handled by the compiler's front end.

\section{Threats to Validity}

\paragraph{Internal Validity.}
The main threat to internal validity (i.e., risk of confounding variables) is that we relied on a source-based heuristic approach to determine the usage of GCC builtins, namely by searching for identifiers of known builtins in the source files.
We could have mistakenly recorded a builtin use when the builtin was enclosed in a comment, or when an identifier with the same name as a builtin was used for another purpose.
However, as described, we used several mitigation strategies to address such ``deceiving'' uses.
Conversely, we could have missed builtin uses if their names consisted of strings that were concatenated by using preprocessor macros; however, we expect such uses to be uncommon.
RQ4 required manual effort for classifying projects based on their use of builtins as they evolved, and selecting representative projects, which is both difficult to reproduce and subjective.
To address this, we had a calibration phrase where three authors performed a classification of 15\% of the projects.

\paragraph{Construct Validity.}
The main threat to construct validity is that the implementation order we suggest might not reflect the needs of developers.
In all implementation scenarios, we assumed that each project has equal weights.
In practice, tool developers might prioritize one project domain over another, and might thus implement builtins in a different order.
For example, developers of security analyzers would likely first want to support projects with a high attack surface, while compilers for embedded systems would implement support only for embedded software.
We consider equal weights as a neutral and simple metric; if not applicable, developers can use our artifact to determine a better-suited order.

\paragraph{External Validity.}
Several threats to external validity (i.e., whether our results are generalizable) are related to the scope of our analyses.
First, besides C code, C++ code also can access GCC builtins, which we considered beyond our scope, so our results cannot be generalized to C++ projects.
We analyzed open-source GitHub projects, hence our findings might not apply to proprietary projects.
Furthermore, they do not necessarily apply to projects hosted on sites other than GitHub; this biases our results as, for example, GNU projects other than GCC are often hosted on Savannah and could potentially rely more strongly on GCC builtins.
Additionally, our results cannot be generalized to the builtins of compilers other than GCC.
Finally, we investigated the usage of builtins at the source level, which might be different from the usage in the compiled binary (e.g., because their usage could be influenced by macro metaprogramming) and the usage during execution of the program.

\section{Related Work}

\paragraph{Studies of inline assembly and linkers.}
Besides compiler builtins, C projects also contain other elements not specified by the C standard.
Rigger et al. found that around 30\% of popular C projects use x86-64 inline assembly~\cite{inlineassembly}.
The current paper demonstrates that GCC builtins are used more frequently than inline assembly, which provides even stronger incentives to implement support by C tools.
Other studies focused on the role of linkers~\cite{linkers} and the preprocessor~\cite{cpreprocessor}.
C projects are often built using Makefiles, whose feature usage has also been investigated~\cite{Martin:make}.

\paragraph{Studies of other language features.}
This paper fits into a recent stream of empirical studies of programming language feature usage, all of which share a methodology of mining software repositories to determine the popularity of features in large sets of open-source projects and/or evaluate the ``harmfulness'' of features in terms of potential for bugs. Most of this work has focused on general-purpose programming languages, and research has evolved from more common to lesser known features.
For example, for Java, the usage of general language features~\cite{dyer:javafeatures,qiu:javafeatures}, fields~\cite{ewan:javafields}, inheritance~\cite{Tempero:javainheritance}, exception handling~\cite{Asaduzzaman:javaexceptions,Nakshatri:javaexceptions,Sena:javaexceptions}, lambda features~\cite{Mazinanian:javalambda} and async constructs on Android~\cite{okur:androidasync} have been studied.
For C++ projects, the usage of templates~\cite{wu:cpptemplates}, generic constructs~\cite{sutton:cppgenerics}, concurrency constructs~\cite{wu:concurrencyconstructs} and asserts~\cite{Casalnuovo:asserts} have been studied. The latter also considered C projects, similar to Nagappan et al.'s study~\cite{Nagappan:goto} of the usage and harmfulness of the goto construct.
However, to the best of our knowledge, a study of the usage of compiler builtins has not yet been conducted, and as such fits into the line of research into C programming language features.

\section{Conclusions}
We have presented an empirical study of the usage of GCC builtins in a corpus of 4,912 open-source C projects retrieved from GitHub.
To the best of our knowledge, this is the first study of compiler builtins despite them having existed in GCC for 30 years.
We believe that they warrant investigation, since more than 12,000 builtins exist that tools could support and because even safety-critical tools such as the CompCert compiler have bugs in the implementations of common builtins.

\paragraph{Implications for tool builders.}
Since \percentageProjectsWithBuiltins{} of all popular projects relied on compiler builtins, any tool that processes C code needs to deal with them.
However, since only about \approxnumberusedbuiltins{} of \nrTotalTerms{} builtins were used, it might not be necessary to implement all builtins.
In fact, we found that architecture-independent builtins are most commonly used and by implementing only \halfprojectnumberbuiltins{} of such core builtins, half of the projects can be supported.
Since the majority of projects mostly added builtin usages, tools are likely still expected to support builtins for code yet to be written.

\paragraph{Implications for the GCC developers.}
We think that our study is also informative for compiler developers, especially those of GCC.
This study demonstrated the large scope of GCC builtin usage and might encourage compiler developers to add, maintain, and document builtins in a consistent and structured way.
In particular, we think that public and internal builtins should be strictly separated.
Since our study highlighted the builtins that remained unused in our data set, such builtins could potentially be considered as deprecation candidates.
As part of our future work, we want to engage with the GCC developers to discuss how the identified problems could be addressed.

\paragraph{Implications for application developers.}
Furthermore, our study informs application developers about downsides of using builtins.
Although supported by mature compilers, GCC builtins are a language extension that are not supported by all tools (e.g., CompCert, TCC, KCC, and Frama-C).
Thus, a reliance on builtins means that fewer code analysis and other tools can be used on such applications.
Furthermore, projects that rely on internal builtins---such as those for variadic arguments handling---add an additional level of technical debt, as they should only be used within GCC and can, in theory, change without notice.
Thus, we recommend application developers to use builtins with caution.

\paragraph{Implications for language designers.}
We believe that our results are also useful to language designers, as they show which functionality plain C lacks, and what potential implications adding compiler builtins has on the projects developed in a given language.

\begin{acks}
We thank the anonymous reviewers for their valuable comments and suggestions to improve the quality of the paper.
We thank Ingrid Abfalter for proofreading an early draft of this paper.
The authors from Johannes Kepler University Linz are funded in part by a research grant from Oracle.
\end{acks}

\balance
\bibliographystyle{ACM-Reference-Format}
\bibliography{paper}

\end{document}